\documentclass[pra,twocolumn,superscriptaddress]{revtex4}
\usepackage{amsmath}
\usepackage[final]{graphicx}
\newcommand{\nc}{\newcommand}
\nc{\bc}{\begin{center}}
\nc{\ec}{\end{center}}

\begin{document}
\title{On the Quantum Phase Operator for Coherent States}
\author{Bo-Sture K.\ Skagerstam}
\email{Bo-Sture.Skagerstam@phys.ntnu.no }
\affiliation{Department of Physics, The Norwegian University of
  Science and Technology,\\ N-7491 Trondheim, Norway }
\affiliation{Microtechnology Center at Chalmers MC2, Department of
  Microelectronics and Nanoscience, \\ Chalmers University of Technology
  and G\"{o}teborg University, S-412 96, G\"{o}teborg, Sweden} 
\author{Bj\o rn \AA. Bergsjordet}
\email{Bjorn.Bergsjordet@phys.ntnu.no}
\affiliation{Department of Physics, The Norwegian University of
  Science and Technology,\\ N-7491 Trondheim, Norway }
\begin{abstract}
  \noindent In papers by  Lynch [Phys. Rev.  {\bf A41}, 2841 (1990)]
  and Gerry and Urbanski [Phys. Rev. {\bf A42}, 662 (1990)] it has
  been argued that the
  phase-fluctuation laser experiments of Gerhardt, B\"uchler and Lifkin
  [Phys. Lett. {\bf 49A}, 119 (1974)]   are in good agreement with
  the variance of the Pegg-Barnett phase operator  for a coherent
  state, even for a small number of photons. We argue
  that this is not conclusive. In fact, we show that the variance of the phase in
 fact
  depends on the relative phase between  the phase of the coherent
  state and  the off-set phase $\phi_0$ of the Pegg-Barnett phase
  operator. This off-set phase is replaced with the phase of a
  reference beam in an actual experiment  and we show that  several choices of
  such a relative phase can be fitted to the experimental data. We also
  discuss the    Noh, Foug\`{e}res and Mandel~[Phys.~Rev.~{\bf A46},
  2840 (1992)] relative phase experiment in
  terms of the Pegg-Barnett phase taking post-selection conditions
  into account.\\ \\PACS Ref:42.50-p;42.50.Gy;42.50.Xa
\end{abstract}
\maketitle

\bc{
\section{INTRODUCTION}
\label{sec:introd}
}\ec

The notion of a quantum phase and a corresponding quantum
phase operator plays an important role in various considerations in e.g.
modern quantum optics (for a general discussion see e.g. Refs.\cite{Carr68,barnett97,SB93}).
Recently it has been argued by R. Lynch~\cite{Lynch90} and C.\ C.\ Gerry
and K.\ E.\ Urbanski~\cite{Gerry90}
that the theoretical values of the variance of the
Pegg-Barnett (PB) phase operator~\cite{Pegg89} evaluated for a
coherent state are in good agreement
with the phase-fluctuation measurements
of Gerhardt, B\"uchler and Lifkin (GBL) \cite{Gerhardt74} for two interfering laser beams. 
In the literature one often finds reiterations of this
statement (see e.g. Ref.\cite{Orzag2000}). For the purpose of
analyzing the experimental data in terms of the PB phase operator one
makes the assumption that the laser light can be described
in terms of a conventional coherent state (see
e.g. Ref.\cite{Klauder&Skagerstam&85}). 
It has, however, been
questioned to what extent this assumption is
correct~\cite{Moelmer1997} based on the fact that
conventional theories of a laser naturally leads to a mixed rather
than a pure quantum state (see e.g. Ref.\cite{Scully&Zubairy} ). Relative
to a reference laser beam the quantum state of the laser can
nevertheless be assumed to be a coherent state \cite{Enk}. We notice  that
the arguments of Ref.\cite{Moelmer1997} has been questioned
\cite{Gea1998}
and that in some laser models
there are indeed mechanisms which may provide for quantum states with
precise values of both the amplitude and the phase. Recent
experimental developments have also actually lead to a precise measurement of the
amplitude and phase of short laser pulses \cite{Walther2001}. 

In the present paper it is assumed
that a coherent state is a convenient
description of the quantum state of the laser in agreement with our
argumentation above. 
Even with the use of coherent states we will claim that a clarification
is required concerning the comparison  between the PB quantum
phase theory and experimental data. 
We will argue that the phase  is naturally given
relative to the PB off-set 
phase $\phi_0$ and that the variance of the relative phase
$\hat\phi-\phi_0$ therefore is dependent on the relative phase between
the phase $\xi$ of
the coherent state and the off-set phase $\phi_0$. In an actual
experiment one measures the phase relative to a reference beam and the
off-set value $\phi_0$ will then effectively be replaced by the phase of the reference
beam.  In the course of our
calculations  and in comparing with experimental data, we will
verify that in some situations the actual phase in the
definition of the  coherent state used is actually irrelevant. In the
course of our considerations below, we will compare the PB approach to
the notion of a quantum phase with other defintions and point out
situations where they are in agreement or disagreement with actual
experimental observations. The paper is organized as follows. In
Section~\ref{sec:pb}  we briefly  review the PB quantum phase operator
theory. Phase fluctuations in the PB theory and in the
Susskind-Glogower (SG) theory \cite{Sussk64,Carr68}  
are discussed in Section~\ref{sec:fluct} and various bounds on phase
fluctuations are derived. Relative PB and SG phase
operators are discussed in Section~\ref{sec:data} together with a
comparison to the GBL experimental data \cite{Gerhardt74}. The PB
theory and the Noh-Foug\`{e}res-Mandel (NFM) \cite{Noh91,Noh92,Noh92_2} operational theory for
a relative phase operator measurement are discussed in Section
\ref{sec:nfm} and, finally, some concluding comments are given in Section \ref{sec:final}.

\vspace{0.5cm}
\bc{
\section{THE PB QUANTUM PHASE OPERATOR}
\label{sec:pb}
}\ec

We make use of a spectral resolution of the PB
phase operator~\cite{Pegg89} defined
on a ($s+1$)-dimensional truncated Hilbert space of states, i.e.
\begin{align}
\label{eq:operator}
  \hat{\phi} & = \sum_{m=0}^s\phi_m \lvert \phi_m\rangle\langle\phi_m\rvert
\end{align}
where
\begin{align}
  \phi_m & = \phi_0 + \frac{2\pi m}{s+1} \ \ \, ; \ \ m=0,1, \ldots, s . 
\end{align}

\noindent In Eq.(\ref{eq:operator}) the normalized state $\lvert \phi_m\rangle $
 can be expressed in terms of the number-operator
eigenstates $\lvert n \rangle$, i.e.
\begin{align}
\lvert \phi_m\rangle = \frac{1}{\sqrt{1+s}}\sum_{n=0}^{s}e^{in\phi_m}\lvert n\rangle~~.
\end{align}
As described by Pegg and Barnett \cite{Pegg89}, we do all the calculations of the
physical quantities in this truncated space and take the limit $s \rightarrow
\infty$ in the end. Care must be taken when performing the appropriate
mathematical limit \cite{Lynch90,Troubles}. Following these definitions, 
the expectation value of a function ${\cal O}$ of the relative phase
operator $\hat{\phi}-\phi_0$ is given by 
\begin{align}
\label{eq:PB}
 \langle {\cal O } \rangle \equiv  \lim_{s\rightarrow \infty} 
\langle \psi \lvert {\cal O}(\hat{\phi}-\phi_0)\rvert\psi\rangle
  &  =  \int_{0}^{2\pi} d\phi \, {\cal O}(\phi) \, P(\phi) ~~,
\end{align}
where $\lvert \psi\rangle$ is a general pure quantum state in the form of a
linear superposition of number-operator eigenstates $\lvert n \rangle$, i.e.
\begin{align}
\label{eq:instate}
\lvert \psi \rangle & = 
\sum_{n=0}^\infty \sqrt{{\cal P}_n}e^{i\xi(n)}\lvert n \rangle~~,
\end{align}
with a normalized number-operator probability distribution ${\cal P}_n$.
Here
\begin{align}
\label{eq:prob}
  P(\phi) & = \frac{1}{2\pi} \left| \sum_{n=0}^\infty
\sqrt{{\cal P}_n}\, 
 e^{in(\phi+\phi_0) - i\xi(n)}\right|^2
\end{align}
is a periodic probability distribution. The distribution $P(\phi)$ is the same
as the one obtained from the SG phase operator
theory \cite{Sussk64},
which has been argued  on general grounds to be the case
\cite{Shapiro91}.   In the case of coherent-like states with
$\xi(n)=n\xi+\xi_0$ but with arbitrary ${\cal P}_n$, the distribution   
$P(\phi)$ depends in general on the
difference between the phase $\xi$ and the PB off-set value $\phi_0$,
i.e. on $\delta\xi \equiv \xi - \phi_0$.  For a coherent state $\lvert
\psi \rangle=\lvert \alpha\rangle$, with $\alpha=
|\alpha|\, e^{i\xi}$, the photon-number distribution is Poissonian,
i.e. ${\cal P}_n = e^{-|\alpha|^2}|\alpha|^{2n}/n!$~. The mean value 
of the number of photons, ${\bar n}$, is then given by ${\bar n}=|\alpha|^2$. In what follows
we will, unless otherwise specified, limit ourselves to the use of
coherent states but our considerations can be extended to
general states, pure or mixed, in a straightforward manner.
%
%
%
%

\vspace{0.5cm}
\bc{
\section{QUANTUM PHASE FLUCTUATIONS}
\label{sec:fluct}
}\ec

We observe that the  variance of the PB phase operator is independent of the off-set phase
$\phi_0$, i.e.
\begin{align}
\label{eq:exp_phi-phi0_square}
\Delta\phi^2 & \equiv  \langle \hat{\phi}^2\rangle
  -\langle \hat{\phi}\rangle^2 = 
  \langle (\hat{\phi}-\phi_0)^2\rangle - \langle
  (\hat{\phi}-\phi_0)\rangle^2~~,  
\end{align}

\noindent but it is dependent on the relative phase
$\delta\xi$, as we will see in detail below.

\begin{figure}[htbp]
  \begin{center}
  \includegraphics[scale=0.44, angle=0]{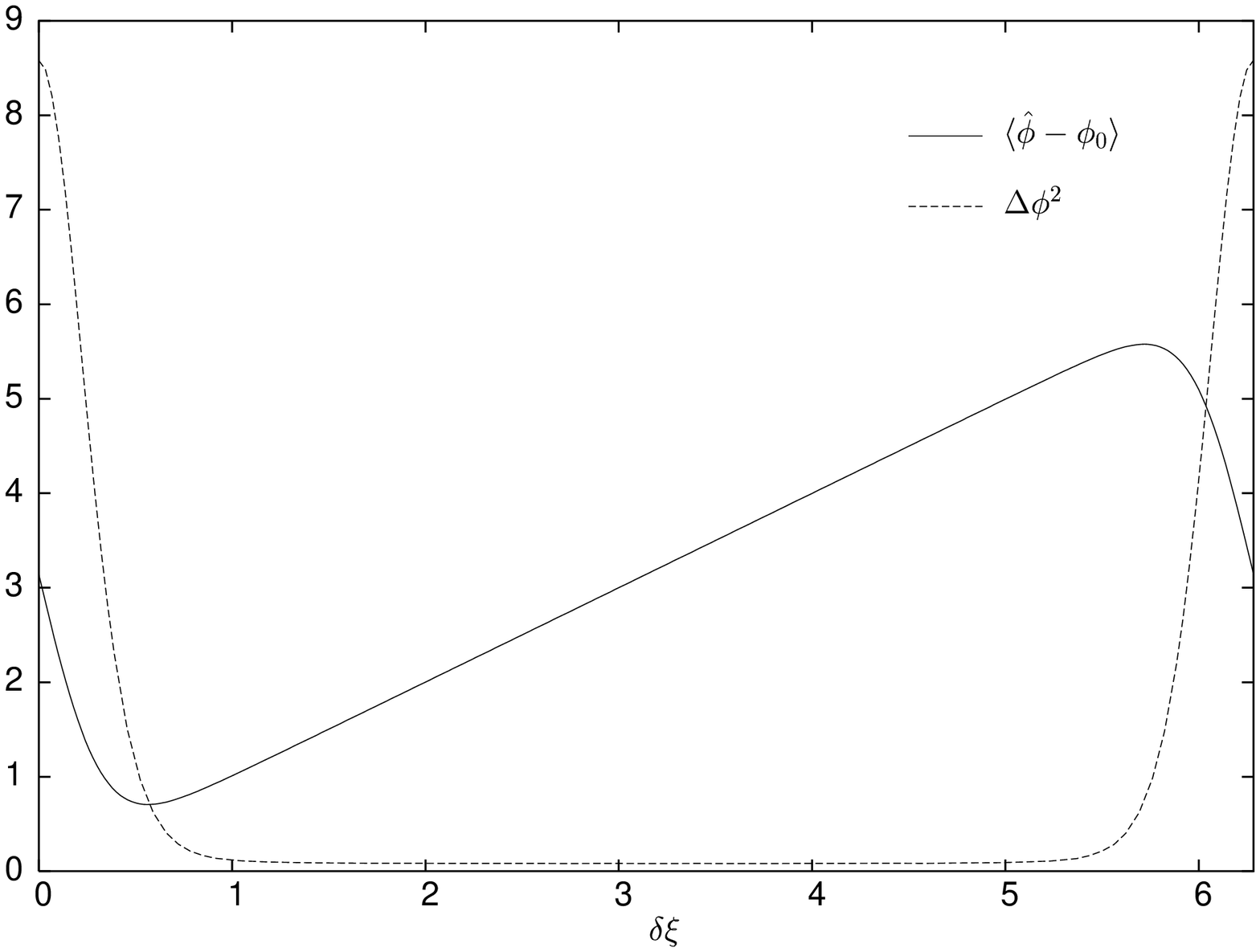}
    \caption{ \label{fig:figure1} The solid line shows the expectation
      value of the relative phase operator $\langle \hat{\phi}
      -\phi_0\rangle$  for different values of the relative phase
      $\delta\xi=\xi -\phi_0$ with $\bar n=4$. The dotted line shows
      the corresponding variance $\Delta \phi^2$. For $\bar n \gg 1$,
      we find that $\langle \hat{\phi} -\phi_0\rangle = \delta\xi$
      except at the boundaries $\delta\xi=0$ or $\delta\xi=2\pi$ where
      $\langle \hat{\phi} -\phi_0\rangle = \pi$ with a maximal
      uncertainty $\Delta\phi = \pi$.}
  \end{center}
\end{figure}

Lower and upper  bounds on the variance $\Delta\phi^2$ can be found as
follows. For a general pure state $|\psi\rangle$ we have
\begin{align}
  \langle [ \hat N,\hat\phi  ]\rangle & =
i(1-2\pi P(0))~~,
 \end{align}
where the distribution $P(\phi)$ is given in Eq.(\ref{eq:prob}) and
a  Heisenberg  uncertainty type of relation follows, i.e.
\begin{align}
\Delta\phi^2\, \Delta N^2 & \geq \frac{1}{4} |1-2\pi P(0)|^2 \ .
\end{align}
For a coherent state, $|\psi\rangle=|\alpha\rangle$, the periodic
distribution $P(\phi)$ 
is now such that the variance $\Delta\phi^2$ has a lower bound
when $\xi-\phi_0= \pm \pi$ (apart from multiples of $2\pi$) with a
mean value of the relative phase operator
$\langle {\hat\phi } - \phi_0 \rangle= \pi$. The minimum value of the variance $\Delta\phi^2$
can then be
found using the same techniques as in the proof \cite{bouten65} of the
implicit bound due to
Judge \cite{judge64}, i.e.
\begin{align}
\Delta N^2\Delta\phi^2 & \geq \frac{1}{4}(1-\frac{3\Delta\phi^2}{\pi^2})^2~~.
\end{align}
From this expression one can easily obtain a lower bound on the
variance 
$\Delta\phi^2$ which we conveniently simplify into the following form
\begin{align}
\label{eq:boundsonvar}
  \frac{1}{4{\bar n}+3/\pi^2}\leq \Delta\phi^2 \leq \pi^2 , 
\end{align}
where we make use of the fact that $\Delta N^2 ={\bar n}$ for a coherent state. 
The lower bound is chosen in such a way that the bound is saturated for the vacuum
distribution with ${\bar n}=0$.
The upper bound is obtained by direct
calculation of the variance using a distribution $P(\phi)$ in the form
\begin{align}
  P(\phi) & = \frac{1}{2}\delta(\phi)+\frac{1}{2}\delta(\phi-2\pi), 
\end{align}
which is valid when the mean number of photons in the coherent state
is such that $\bar n \gg 1$ and $\delta\xi = 0$. 
%

In Fig.~\ref{fig:figure1} we show the expectation value of the relative phase
operator ${\hat \phi}-\phi_0$ and the corresponding variance $\Delta\phi^2$
for a coherent
state with a mean number of photons ${\bar n}=|\alpha|^2=4$ as a function of the
relative phase $\delta\xi$ of the coherent state, which due to 
the periodicity of $P(\phi)$ always can be
chosen in the same range as $\phi$.
The expectation value and the variance 
are periodic functions of the variable  $\delta\xi$. When ${\bar n}$ is increased
$\Delta\phi^2$ becomes more narrow around the values 
 $\delta\xi= 0$ and $\delta\xi=2\pi$. Except for these
 boundary points  $\langle \hat\phi -\phi_0 \rangle $ approaches the expected linear
 dependence of $\delta\xi$. The PB phase operator theory therefore
 predicts a small $\Delta\phi^2$ for ${\bar n}\gg 1$ except
 for unavoidable periodic spikes with $\Delta\phi^2= \pi^2$.

\begin{figure}[htbp]
\begin{center}
%
    \includegraphics[scale=0.350, angle=90]{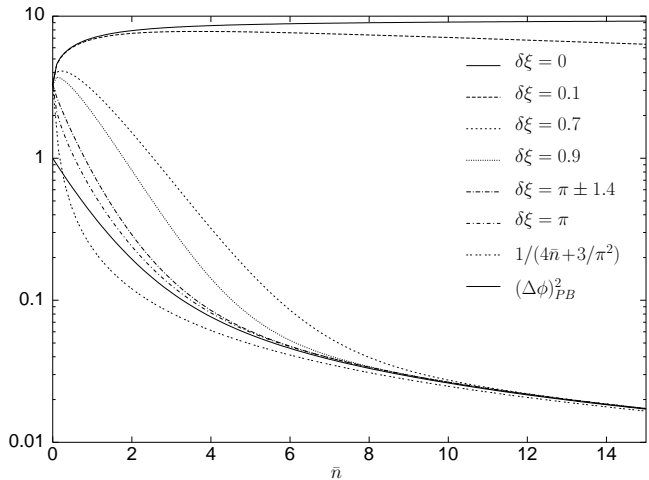}
    \caption{ \label{fig:figure2} This figure illustrates the variance
      $\Delta\phi^2$ and
its dependence on the relative phase difference
$\delta\xi=\xi-\phi_0$. The variance is plotted as a function of the
mean number of photons in the coherent state, i.e. $\bar
n=|\alpha|^2$. Lower (lowest dotted curve)  and upper bounds (upper solid curve)
on $\Delta\phi^2$ in  accordance with  Eq.(\ref{eq:boundsonvar}) as well as
the variance $(\Delta\phi)^2_{PB}$ (lower solid curve), as defined in terms
of cosine and sine phase operators according to  Eq.(\ref{eq:pbdef1}), 
are also shown.  The variance $(\Delta\phi)^2_{PB}$ is independent of 
$\delta\xi$.}
\end{center}
\end{figure}

In Fig.~\ref{fig:figure2} we illustrate how the variance $\Delta\phi^2$
depends on the relative phase difference $\delta\xi$ as a function
of the mean number ${\bar n}$ of photons of the coherent 
state  together with the upper and lower bounds in accordance with
Eq.(\ref{eq:boundsonvar}).
As is seen from Eqs.~(\ref{eq:PB})
and~(\ref{eq:exp_phi-phi0_square}) the variance $\Delta \phi^2$ is symmetric around
$\delta\xi=\pi$. If $\delta\xi$ is a multiple of $2\pi$, we find that
$\Delta \phi^2$  approaches its
maximum value $\pi^2$ fast as ${\bar n}\rightarrow \infty$. For all other values of  $\delta\xi$ we
find that  $\Delta \phi^2$
approaches $1/4{\bar n}$ if ${\bar n}$ is large enough.
In Fig.~\ref{fig:figure2} we also show 
the variance
$(\Delta\phi)^2_{PB}$ as expressed in terms of cosine and sine phase
operators as used in SG-theory \cite{Sussk64,Carr68}. 
The variance
$(\Delta\phi)^2_{PB}$ is then evaluated in terms of the PB phase
operator $\hat\phi$
according to
\begin{align}
\label{eq:pbdef1}
  (\Delta\phi)^2_{PB}& \equiv  (\Delta\cos(\hat\phi -\phi_0))^2 +  
(\Delta\sin(\hat\phi -\phi_0))^2~~,
\end{align}
where we in general define
\begin{align}
  (\Delta f(\hat\phi - \phi_0))^2 & \equiv \langle f^2(\hat\phi-\phi_0) \rangle
  - \langle f(\hat\phi - \phi_0) \rangle^2~~.
\end{align}
A straightforward calculation  making use of the distribution
Eq.(\ref{eq:prob}) then leads to the result
\begin{align}
\label{eq:pbd1}
  (\Delta\phi)^2_{PB}& =1- \left[\psi_{PB}(\bar n)\right]^2 \ , \\
\psi_{PB} (\bar n) & = \sqrt{\bar n} e^{-\bar n} \sum_{n=0}^\infty
  \frac{\bar n^n}{\sqrt{n!(n+1)!}} \ .
\end{align}

\noindent In obtaining this expression we have made use of the relation
\begin{align}
\label{eq:trigo}
 \langle e^{i({\hat \phi} - \phi_0 )}\rangle = e^{i\delta\xi}\psi_{PB} (\bar n)~~,
\end{align}
which shows that for elementary trigonometric functions the PB phase
 operator for a coherent state only leads to a modified amplitude for a small average
 number ${\bar n}$. If we define the exponential $e^{i{\hat \phi}} =
 {\hat C} + i{\hat S}$ in terms
 of the SG Cos- and Sin-operators ${\hat C}$ and ${\hat S}$ \cite{Sussk64,Carr68},
 the SG theory also leads to Eq.(\ref{eq:trigo}) apart from the
 $\phi_0$ dependence. The corresponding expression for the
 fluctuations $ (\Delta\phi)^2_{SG}$ in the SG-theory  follows from the results of
 Ref.\cite{Carr68}, i.e.
\begin{align}
 (\Delta\phi)^2_{SG} \equiv  \langle {\hat C}^2 + {\hat S}^2 \rangle
 - \langle {\hat C} \rangle ^2  - \langle {\hat S} \rangle ^2=  (\Delta\phi)^2_{PB} -
 \frac{1}{2}e^{- \bar n}~~,
\end{align}
where $\langle \cdot \rangle$ denotes a conventional quantum-mechanical expectation value.
We notice that the fluctuations
$(\Delta\phi)^2_{PB}$ and  $(\Delta\phi)^2_{SG}$  do not depend on the phase $\xi$. This
 independence of the phase $\xi$ does not imply that this is an unessential
 parameter. The coherence property of the pure state as given by
 Eq.(\ref{eq:instate}) is essential in
 obtaining the result Eq.(\ref{eq:pbd1}). If we instead consider a mixed
 state as described by the diagonal density matrix 
$\rho = \sum_{n=0}^{\infty}  {\cal P}_n \lvert n \rangle \langle n
 \rvert $ we would e.g. obtain the results
\begin{align}
 \langle \hat\phi -\phi_0\rangle = \pi~~,~~
\Delta\phi^2 = \frac{\pi^2}{3}~~,~~(\Delta\phi)^2_{PB} = 1~~,
\end{align}
and
\begin{align}
 1 \geq (\Delta\phi)^2_{SG} = 1 - \frac{1}{2}{\cal P}_0 \geq \frac{1}{2}~~.
\end{align}

\noindent The explicit
result Eq.(\ref{eq:pbd1}) can be used to derive the following
convenient upper and lower
bounds
\begin{align}
  \frac{1}{1+4{\bar n}}\leq (\Delta\phi)^2_{PB}\leq 1 ~~.
\end{align}
\begin{figure}[htbp]
\begin{picture}(200,100)(50,90)
%
    \includegraphics[scale=0.350, angle=90]{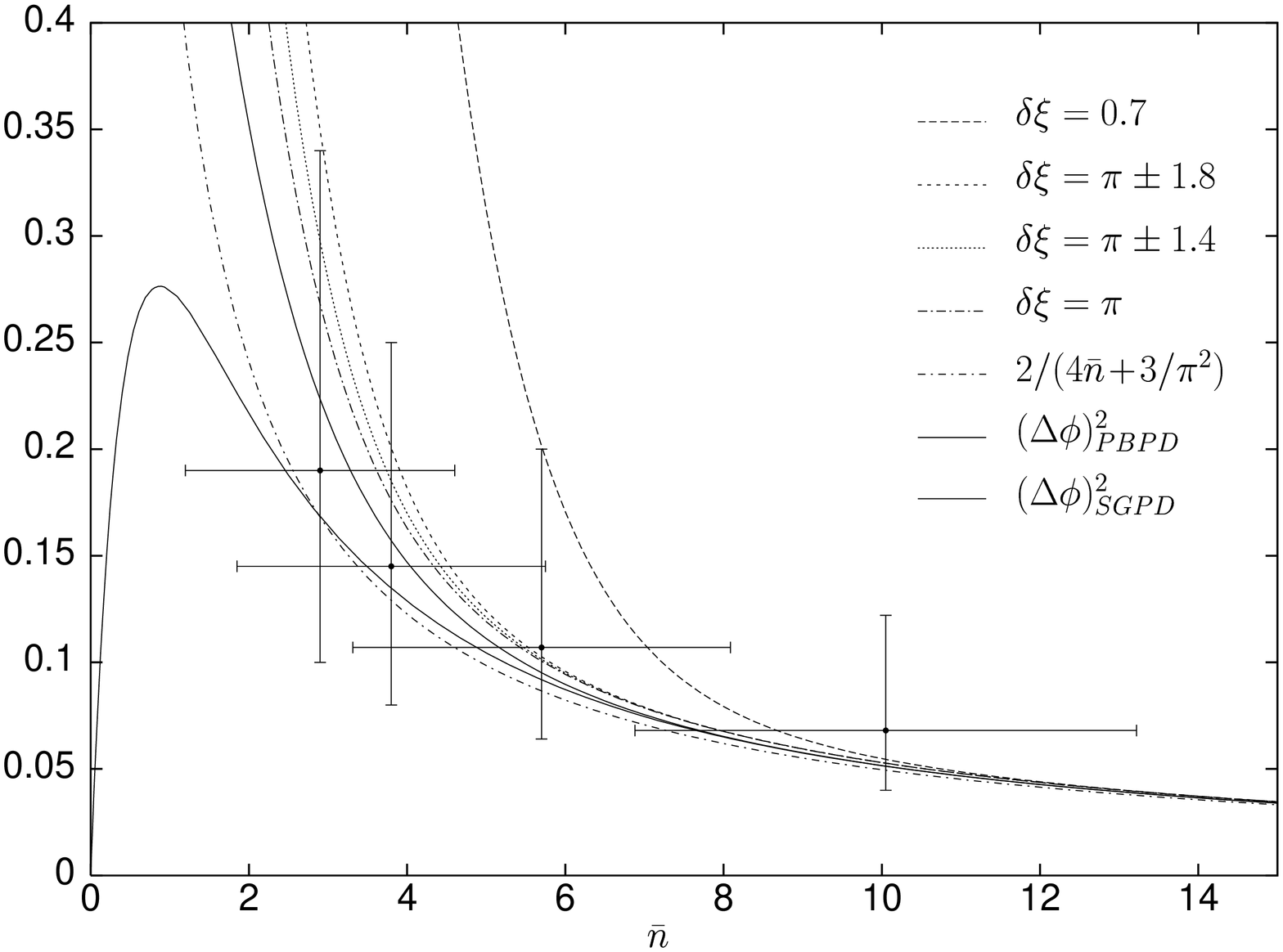}
\end{picture}
\vspace{30mm}
    \caption{ \label{fig:figure3}  This figure compares the $\delta\xi$-dependent
      variance $2\Delta\phi^2$ as a function of the mean
      number of photons in the coherent state for different values of
      the relative phase     $\delta\xi=\xi -\phi_0$ as well as the Susskind-Glogower (lower
      solid curve) and the
      Pegg-Barnett (upper solid curve) relative phase fluctuations
      $\Delta\phi^2_{SGPD}$ 
and  $\Delta\phi^2_{PBPD}$
      according to Eqs.(\ref{eq:sgdiff}) and  (\ref{eq:pbdiff}) 
 respectively, which do not depend on $\xi$, and 
      the GBL data from Ref.\cite{Gerhardt74}. In addition horizontal
      error-bars of width $\bar n^{1/2}$ are added to the GBL data.} 
\end{figure}
%

\bc{
\section{RELATIVE QUANTUM PHASE OPERATORS AND COMPARISON TO THE
  GBL-DATA}
\label{sec:data}
}\ec

In Ref.~\cite{Gerhardt74} one has measured phase fluctuations of two
interfering laser beams. A comparison of these experimental data with the
fluctuations of the 
relative SG phase operator as given by
\begin{align}
\label{eq:sgdiff}
  (\Delta\phi)^2_{SGPD} = 1-e^{-\bar n}- \left[\psi(\bar n)\right]^2 ~~, 
\end{align}
where
\begin{align}
\psi (\bar n) = \bar n e^{-2\bar n} \left[ \sum_{n=0}^\infty
  \frac{\bar n^n}{n!\, (n+1)^{1/2}} \right]^2 = \psi_{PB}({\bar n})^2 ~~,
\end{align}
was discussed in great detail in Ref.~\cite{Nieto77}. Here we observe that
$(\Delta\phi)^2_{SGPD}$ does not depend on the phase $\xi$.
The 
 experimental data of Ref.\cite{Gerhardt74} (GBL-data) used in
the figures of this article are listed in~Ref.\cite{Nieto77}. To these
data we have added horizontal error bars of width $\bar n^{1/2}$. In
Fig.~\ref{fig:figure3} we plot the  GBL-data and
$(\Delta\phi)^2_{SGPD}$. Since the GBL-data actually corresponds to two
separate and independent measurements of phase fluctuations we also
compare the GBL-data with the PB phase fluctuations
$2(\Delta\phi)^2$. In analogy with the fluctuations of the relative SG phase operator 
it is of interest also to compare these
experimental data with fluctuations of the relative PB phase operator as defined by
\begin{align}
\label{eq:pbdiff}
  (\Delta\phi)^2_{PBPD}& \equiv  (\Delta\cos(\hat\phi_1 -\hat\phi_2))^2 +  
(\Delta\sin(\hat\phi_1 - \hat\phi_2))^2~~,
\end{align}
extending Eq.(\ref{eq:pbdef1}) to two independent phase measurements with PB phase
operators $\hat\phi_1$ and $\hat\phi_2$ with a joint distribution
$P(\phi_1,\phi_2)= P(\phi_1)P(\phi_2)$. The distributions $P(\phi_1)$ and $P(\phi_2)$
are then assumed to be equal, apart from the dependence of a possible
optical path length difference  which will not effect our results in
the end. 
A straightforward calculation leads to
the result
\begin{align}
 (\Delta\phi)^2_{PBPD}= 1-(\psi_{PB}({\bar n}))^4~~.
\end{align}
It appears from Fig.~\ref{fig:figure3} that $ (\Delta\phi)^2_{PBPD}$
provides the best fit to the GBL data. In view of the fact that 
$ (\Delta\phi)^2_{PBPD}$ does not depend on any optical path difference
or on the phase $\xi$ suggest to us that this measure of phase
fluctuations is appropriate at least as far as the GBL data is
concerned. As far as we can see, these  results are not in complete
agreement with those presented by Lynch (1995)\cite{Lynch90}. 

\begin{figure}[htbp]
  \begin{center}
  \includegraphics[scale=0.5, angle=0]{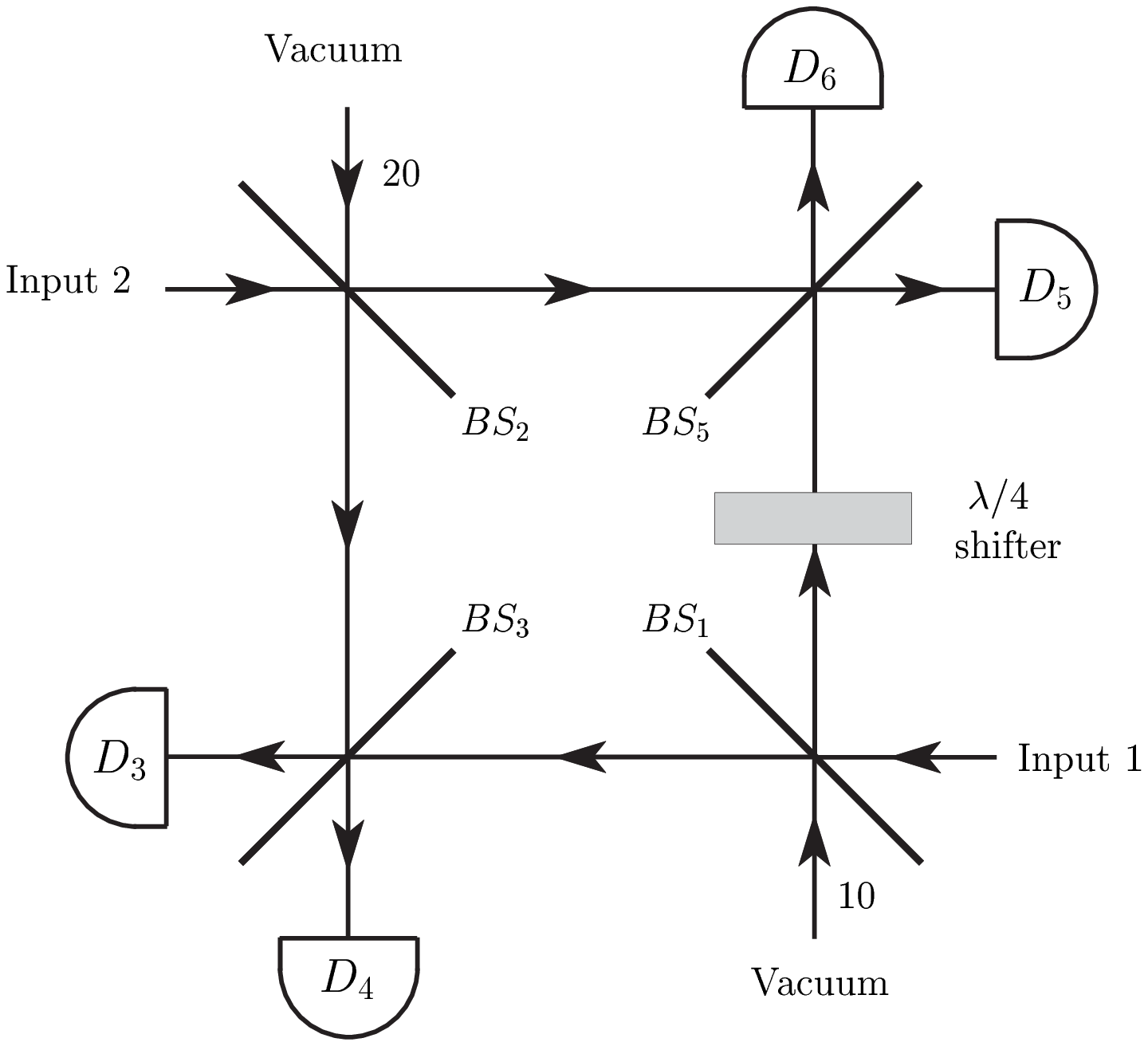}
    \caption{ \label{fig:figure4} The experimental setup as described
      by Noh,  Foug\`{e}res  and Mandel 
      in Refs.\cite{Noh91,Noh92}. 
     }
  \end{center}
\end{figure}

\begin{figure}[htbp]
  \begin{center}
  \includegraphics[scale=0.4, angle=0]{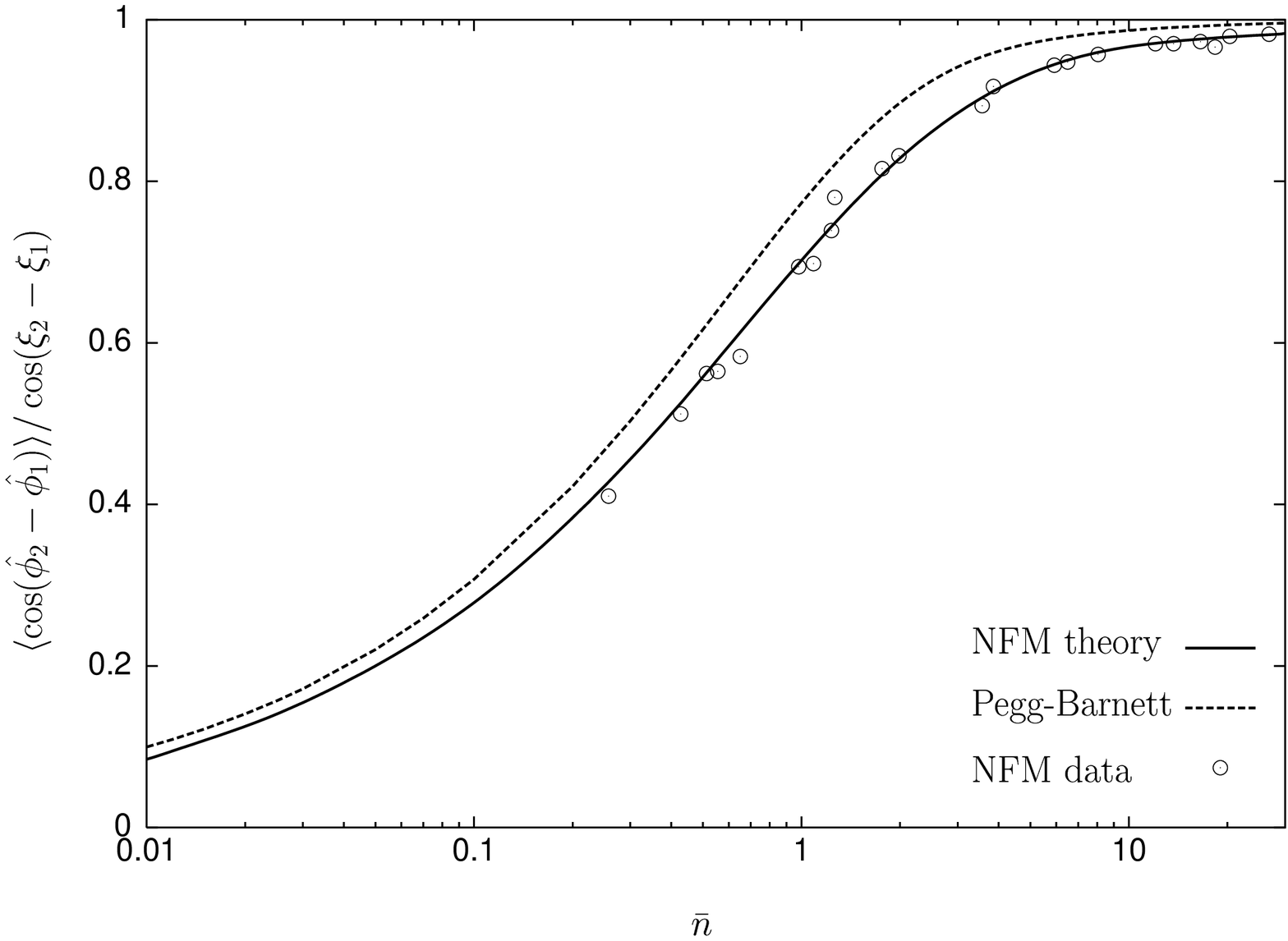}
    \caption{ \label{fig:figure5} The expectation value
      $\langle\cos(\hat\phi_2
      -\hat\phi_1)\rangle/\cos(\xi_2-\xi_1)\approx \psi_{PB}({\bar n})$ according to the PB
      theory as a function of the average
      number of photons in the input port 1  using the experimental
      setup of Fig.~\ref{fig:figure4}.  The input port 2 has a mean value
      ${\bar n}_2=50$ of photons. The SG-theory gives the same
      result which does not depend on the phase difference $\xi_2 -\xi_1$.
The NFM theory and NFM
    data curves are read from similar figures in Ref.\cite{Noh92}.}
  \end{center}
\end{figure}

\begin{figure}[htbp]
  \begin{center}
  \includegraphics[scale=0.35, angle=90]{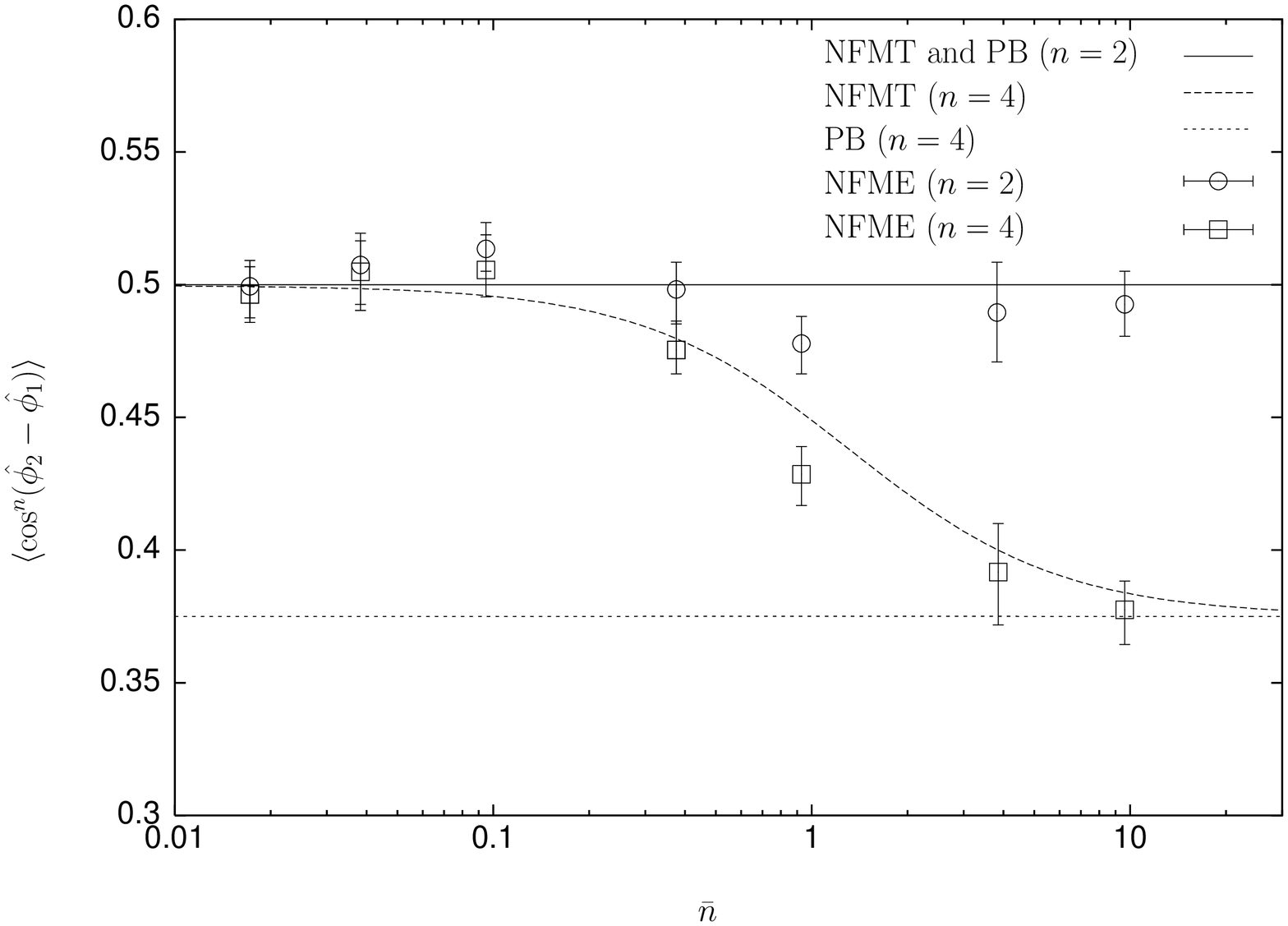}
    \caption{ \label{fig:figure7} The figure compares the expectation
      values $\langle\cos^n(\phi-\phi_0)\rangle$ evaluated with the PB theory
      and the  NFM  theory (NFMT)  and experimental data (NFME) for
      $n=2$ and $n=4$ as a function of the number of quanta in
      input port 2
 ($\bar n = |\alpha |^2$)  according to Fig.~\ref{fig:figure4} with a
      vacuum for input port 1.}
  \end{center}
\end{figure}

\begin{figure}[htbp]
  \begin{center}
  \includegraphics[scale=0.35, angle=90]{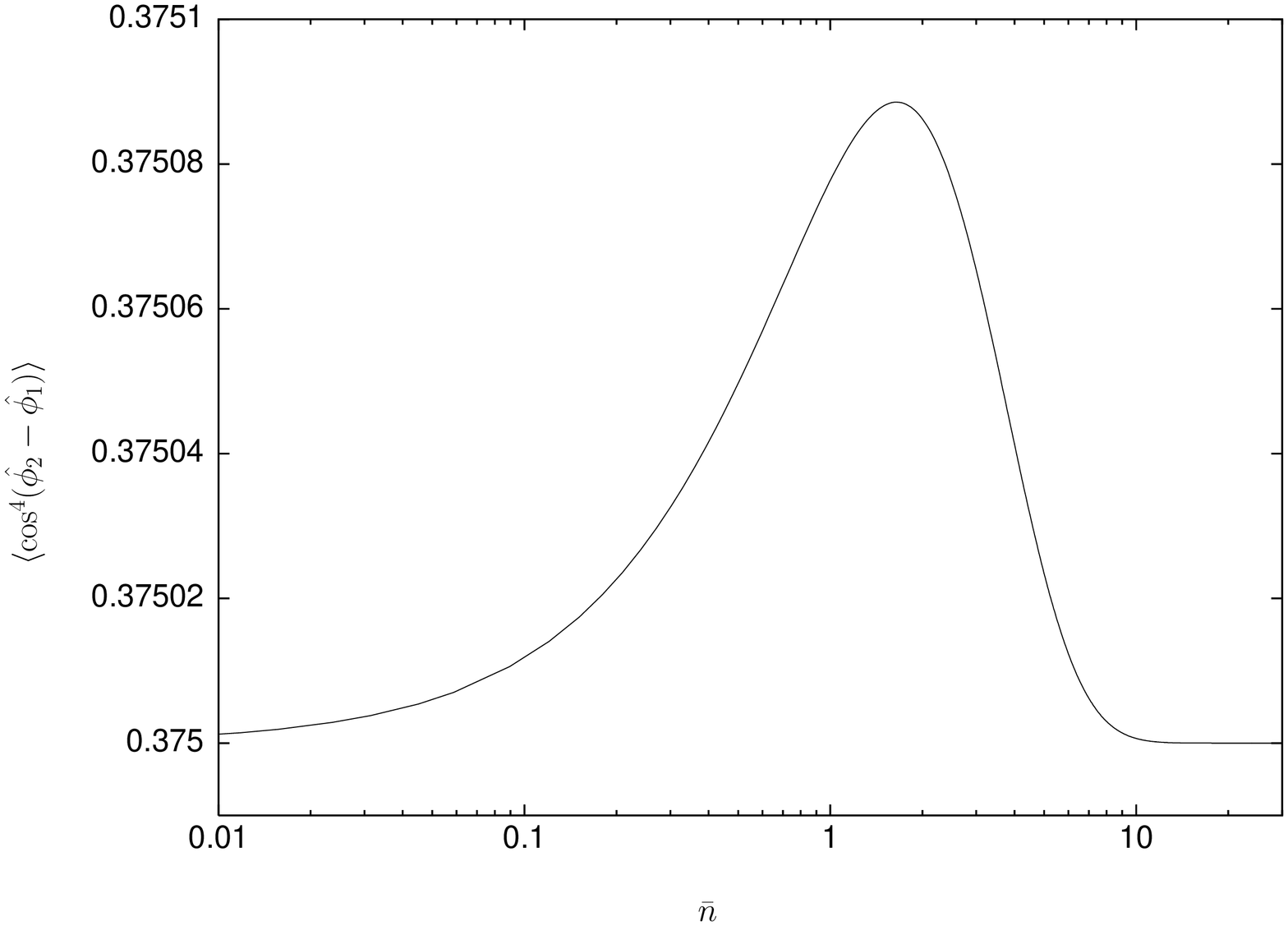}
    \caption{\label{fig:figure8} The expectation value
      $\langle\cos^4(\phi-\phi_0)\rangle$ evaluated with the PB theory
as a  function of
      $\bar n=|\alpha|^2$ with the NFM post-selection condition
      included as described in the main text. The experimental setup
      is as in Fig.~\ref{fig:figure7}.} 
  \end{center}
\end{figure}

\begin{figure}[htbp]
  \begin{center}
  \includegraphics[scale=0.40, angle=90]{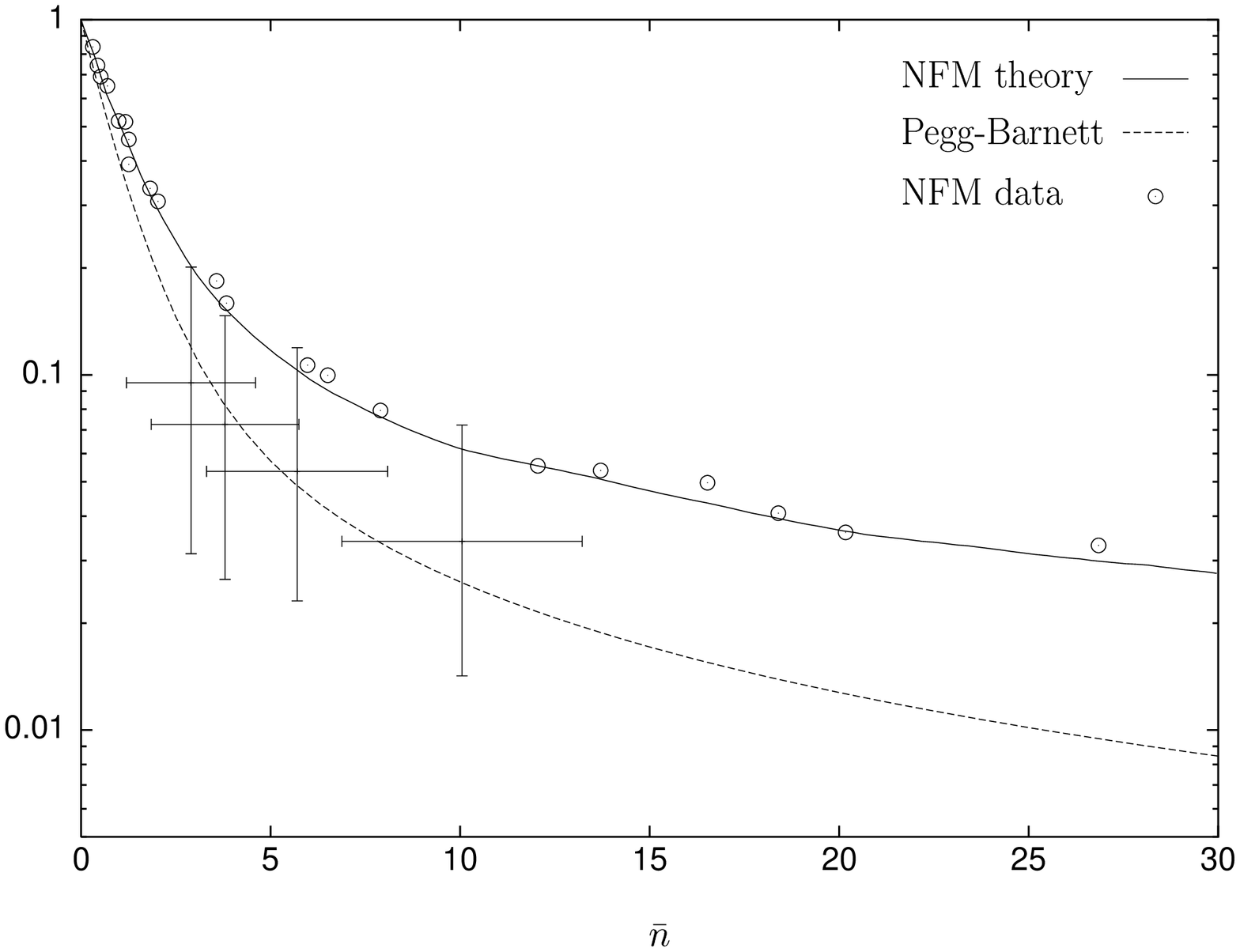}
    \caption{ \label{fig:figure6} This figure compares the relative
      phase variance
    $(\Delta\cos({\hat \phi_1}-{\hat \phi_2}))^2
+(\Delta\sin({\hat \phi_1}-{\hat \phi_2}))^2$
      evaluated with the PB theory and 
the corresponding  NFM theory variance $\langle(\Delta \hat C_M)^2\rangle
    + \langle(\Delta \hat S_M)^2\rangle $, the NFM data of
    Ref.\cite{Noh92} and the GBL data from Ref.\cite{Gerhardt74}. 
The conditions are as described in
      Fig.~\ref{fig:figure5}.
In addition horizontal
      error-bars of width $\bar n^{1/2}$ are added to the GBL data.}
  \end{center}
\end{figure}

\vspace{0.5cm}
\bc{
\section{THE NFM OPERATIONAL APPROACH AND COMPARISON WITH THE PB
  THEORY}
\label{sec:nfm}
}\ec
In Refs.\cite{Noh91,Noh92,Noh92_2} a new formalism~(NFM) for the phase difference
between the states of two quantized electromagnetic fields is explored both
theoretically and experimentally. The experimental setup is
illustrated in Figure~\ref{fig:figure4}. In their experiments 
the relative phase is determined by counting the number of photons 
detected in each detector within a time interval, disregarding
measurements when the number of photons in detectors $D_3$ and $D_4$
and  detectors $D_5$ and $D_6$ are equal. The experimental accuracy is
considerably increased as compared to the GBL results. As illustrated in
 e.g. Figs.~\ref{fig:figure5}-\ref{fig:figure6}   the inclusion of
error bars for the NFM experimental data would barely be visible. 

Here we reconsider the calculation of some functions of the relative 
phase operator ${\hat \phi_2}-{\hat \phi_1}$ 
by making use
of the PB-theory taking into account the post-selection mentioned above, i.e. disregarding
measurements when the number of photons in detectors $D_3$ and $D_4$
and  detectors $D_5$ and $D_6$ are equal.
We therefore calculate all the expectation values within the PB scheme, 
by first evaluating the complete expectation value $\langle
\cos^n(\hat\phi_2 -\hat\phi_1)\rangle$ according to
Eq.~(\ref{eq:PB}) extended to two independent PB phase operators. We 
then subtract the contributions discarded by NFM in their
experiment, i.e.
\begin{eqnarray}
\lim_{s\rightarrow\infty}
\sum_{m_3=0}^s \sum_{m_4=0}^s \sum_{m_5=0}^s \sum_{m_6=0}^s 
\langle
m_3|\langle m_4| \langle m_5 | \langle m_6|\,  \rho \cdot \, \nonumber \\ \cos^n(\hat
\phi_2 -\hat\phi_1)
|m_6\rangle |m_5 \rangle | m_4\rangle |m_3 \rangle \delta_{m_3,m_4}\delta_{m_5,m_6}
\end{eqnarray}
and renormalize the final result with the factor \cite{Noh91,Noh92,Noh92_2}
\begin{align}
\label{eq:renorm}
   N & =  1-e^{-(|\alpha_1|^2+|\alpha_2|^2)} \, 
 I_0\left(\frac{|\alpha_1^2 - \alpha_2^2|}{2}\right) \,
 I_0\left(\frac{|\alpha_1^2 + \alpha_2^2|}{2}\right) \ ,
\end{align}
where $I_0$ denotes a modified Bessel function. 
The initial density matrix $\rho$ has been assumed to be given by
\begin{align}
  \rho & = |\alpha_1\rangle|\alpha_2\rangle |0\rangle_1|0\rangle_2 \,
  \langle 0|_2\langle 0|_1\langle \alpha_2| \langle \alpha_1|  \ ,
\end{align}
where the indices to the vacuum state indicates vacuum port 1 and 2
according to Figure~\ref{fig:figure4}. The normalizing factor $N$
as given in Eq.(\ref{eq:renorm}) is obtained by calculating the trace of this initial
density matrix taking the post-selection condition into account.
Input port 1 and 2 are in the
coherent states $|\alpha_1\rangle$ and $|\alpha_2\rangle$ respectively
with $\alpha_1=|\alpha_1|e^{i\xi_1}$ and
$\alpha_2=|\alpha_2|e^{i\xi_2}$. 

In Fig.~\ref{fig:figure5} we present the result of the calculation of
$\langle\cos({\hat \phi_2}-{\hat \phi_1})\rangle/\cos(\xi_2 - \xi_1)$
as a function of the average number of photons in port 1, i.e.
 ${\bar n}\equiv {\bar n}_1= |\alpha_1|^2$, for a fixed large average number of photons
 in port 2 ( ${\bar n}_2= |\alpha_2|^2 = 50$). 
As discussed in Refs.\cite{Noh91,Noh92,Noh92_2} 
the averages ${\bar n}_1$ and ${\bar n}_2$ should be replaced by the observed averages taking
 the experimental detection efficiency into account. Since ${\bar n}_2$
 is large, the post-selection
 restriction above can be disregarded with an exponentially small
 error. Furthermore, the observable ${\hat \phi_2}$ can for
 sufficiently large ${\bar n}_2$
 be replaced by $\xi_2$ and a straightforward calculation then leads
 to
\begin{align}
\langle\cos({\hat \phi_2}-{\hat \phi_1})\rangle/\cos(\xi_2 - \xi_1)=
\psi_{PB}({\bar n})~~,
\end{align}
independent of $\xi_2 - \xi_1$. We find that the PB-theory, which in
this case agrees with the SG-theory,  predicts results which lie
above the experimental data as presented in Fig.~\ref{fig:figure5}. On
this issue we are not in agreement with Ref.\cite{Noh92_2} since their corresponding
curve lies below the experimental data. Our conclusion is, however,
the same: due to the small error-bars the PB-theory does not agree
with NFM experimental data in this case.

We now consider other observables considered in
Refs.\cite{Noh91,Noh92,Noh92_2} but which were not calculated using
the PB-theory.
The  expectation value  $\langle
\cos^2(\hat\phi_1-\hat\phi_2)\rangle$ with the setup as given 
by Fig.~\ref{fig:figure4},
where the input port 2 is in a coherent state $|\alpha\rangle$ and the
input port 1 is the
vacuum field, is e.g. given by
\begin{align}
  \langle\cos^2(\hat\phi_2 -\hat\phi_1)\rangle & = \frac{1}{2}~~,
\end{align}
since the distribution $P(\phi_1)$ for the observable 
${\hat \phi_1}$ in this case is a constant and the averaging of the
  observable ${\hat \phi_2}$ with the post-selection of
  Fig.~\ref{fig:figure4} leads to the normalization factor as given 
by Eq.(\ref{eq:renorm}).
This result agrees exactly with the NFM theory and also with
the experimental data
as seen in Fig.~\ref{fig:figure7}. Even though the probability
distributions of the relative phase in the PB and the NFM theory
has been argued to be different in the NFM experimental
situation\cite{Noh93},
  some observables can nevertheless apparently lead
to the same result. In a similar calculation of the corresponding
expression using the SG-theory
\cite{Carr68} we replace the 
$\cos^2(\hat\phi_2 -\hat\phi_1)$ PB operator by the square of the operator
\begin{align}
\label{eq:C12}
{\hat C}_{12}= {\hat C}_{1}{\hat C}_{2}+ {\hat S}_{1} {\hat S}_{2}~~,
\end{align}
where  the SG-theory operators ${\hat C}_{k}$ and ${\hat S}_{k}$ corresponds, for
$k=1,2$, to
the PG-theory operators $\cos(\hat\phi_k)$ and $\sin(\hat\phi_k)$
respectively. 
A calculation of $\langle {\hat C}_{12}^2 \rangle$, making use of the definition
Eq.(\ref{eq:C12}) and
with the conditions of Fig.~\ref{fig:figure7} for a sufficiently
large mean value ${\bar n}_2$ of photons in input port 2, i.e. when one can
disregard effects of the NFM post-selection restriction mentioned
above, 
then leads to the result
\begin{align}
\langle {\hat C}_{12}^2 \rangle  = \frac{1}{4}(1-e^{-{\bar n}})~~,
\end{align}
with  an asymptotic value of $1/4$. As seen from Fig.~\ref{fig:figure7} this asymptotic value is
not in agreement with the NFM experimental data.

The general expression of the expectation value  $\langle
\cos^4(\hat\phi_2 -\hat\phi_1)\rangle$ for  the experimental setup as given by 
Fig.~\ref{fig:figure4},
where the input port 2 again  is in a coherent state $|\alpha\rangle$
and the input port 1 is in the vacuum field, is given by
\begin{align}
  \langle \cos^4(\hat\phi_2 -\hat\phi_1)\rangle & = \frac{3}{8} -
  \frac{T}{N}
\end{align}
where
\begin{align} 
T & = \frac{3}{2}
  e^{-|\alpha|^2}\left\{-\frac{1}{12288}|\alpha|^4 + A + B\right\}
\end{align}
with
\begin{eqnarray}
A =   \frac{1}{4}\sum_{m5=0}^\infty (\frac{1}{4}|\alpha|)^{2(m_5+3)}\frac{1}{(m_5+1)!^2} \cdot
 \nonumber \\
 \frac{m_5^2+2m_5-2}{\sqrt{{6(2m_5+3)(m_5+2)^3(2m_5+5)(m_5+3)^3}}}
\end{eqnarray}
and
\begin{eqnarray}
B = ~~~~~~~~~~~~~~~~~~~~~~~~~~~~~~~~~~~~~~\nonumber \\
\frac{1}{8}\sum_{m3=0}^\infty\sum_{m5=0}^\infty
  (\frac{1}{4}|\alpha|)^{2(m_3+m_5+4)}
  \frac{1}{(m_3+2)!^2(m_5+2)!^2} \cdot \nonumber \\
\frac{(m_3+m_5+4)(m_3+m_5+3)-4(m_3+2)(m_5+2)}{\sqrt{6(2m_3+2m_5+5)(m_3+m_5+3)}}\cdot 
\nonumber \\
\frac{1}{\sqrt{(2m_3+2m_5+7)(m_3+m_5+4)}}~.~~~~~~~~
\end{eqnarray}

\noindent A very accurate analytical  approximation of this  expression is
\begin{eqnarray}
  \langle \cos^4(\hat\phi_2 -\hat\phi_1)\rangle \approx ~~~~~~~~~~~~~~~~~~~~~~~~~\, \nonumber \\
  \frac{3}{8}+\frac{3}{2}\, e^{-|\alpha|^2} \,
\left({\frac{1}{12288}|\alpha|^4
  +\frac{\sqrt{15}}{4423680}|\alpha|^6}\right)/N~~~.
\end{eqnarray}
For small values of $|\alpha|^2$, ($\, |\alpha|^2\leq 1$), we also
find that 
\begin{align}
  \langle\cos^4(\hat\phi_2 -\hat\phi_1)\rangle \approx \frac{3}{8} +
  \frac{1}{8192}|\alpha|^2+\frac{1}{65546}(\frac{\sqrt{15}}{45}-3)|\alpha|^4~~~,
\end{align}
is an accurate analytical approximation with an error of less than $1\%$.

In Fig.~\ref{fig:figure7} we compare the expectation values
$\langle\cos^n(\hat\phi_2 - \hat\phi_1)\rangle$ with $n=2$ and $n=4$
of the PB theory  with
NFM data and theory. As we noticed above, for $n=2$ the curves overlap, but
with $n=4$ the curves are completely different. The theoretical PB curve for
$n=4$ is actually  very close to the constant $\frac{3}{8}$ for all values
of ${\bar n}$. The effect of the post-selection is not visible
in Fig.~\ref{fig:figure7}. In 
Fig.~\ref{fig:figure8} we have enlarged the portion of
Fig.~\ref{fig:figure7} where the post-selection is important and it is
seen that the NFM post-selection only leads to a very small numerical
correction for ${\bar n}\leq 10$.

As we see in Fig.~\ref{fig:figure6} the values of phase
fluctuations found
from the PB theory 
are in good agreement with the experimental results of GBL. We also
notice that NFM data and theory lie at the edge of the accepted
variance of the GBL data. The GBL experimental data have here been adjusted to apply to 
the experimental setup as presented in
Fig.~\ref{fig:figure4}. In contrast to the GBL experimental procedure
we now do not have two independent
measurements. The necessary adjustments are a division of $2$ of  the
GBL data 
and a corresponding division by $\sqrt{2}$ of  the
variances as quoted by GBL \cite{Gerhardt74}. 
%
The NFM experimental data  as well as the NFM theoretical  values used in the figures of the
present paper are read from the corresponding
figures in the article~\cite{Noh92} by 
importing the relevant figures and  making use of a graphical and computer-based
numerical routine with a sufficient numerical accuracy.

\vspace{-0.5cm}
\bc{
\section{FINAL REMARKS}
\label{sec:final}
}\ec
\vspace{-0.5cm}

In summary, we have reconsidered some aspects of quantum operator
phase theories and recalculated various expectation values of relative phase
operators using in particular the PB-theory and,  with regard to the NFM experimental
data, we have taken appropiate post-selection constraints into account
when required. We have
also  considered a set of observables which has been measured but not
previously calculated using the PB-theory.  We have seen that there are
definitions of phase fluctuations which do not depend on the actual
phases of coherent states used to describe the quantum states to be probed,
even though the purity of the states are important. The PB-theory appears to describe
accurately some experimental data but not all. Some of our
results are in
 disagreement with similar results
available in the literature but we, nevertheless, reach a similar conclusion as in
the NFM theory \cite{Noh91,Noh92,Noh92_2}, i.e. the notion of a relative
quantum phase depends on the
actual experimental setup.  We have limited our
considerations to the GBL~\cite{Gerhardt74}- and the NFM~\cite{Noh91,Noh92,Noh92_2}- 
experimental data. Further experimental considerations has been
discussed in e.g. Ref.\cite{Torgerson96}, and commented upon in
Ref.\cite{Font02}, 
illustrating again
that
the notion of a relative quantum phase appears to depend on the
experimental situation.
$\\\\\\$

\begin{center}
{\bf ACKNOWLEDGMENT}
\end{center}
%

One of the authors (B.-S.S.)  wishes to thank NorFA for financial support and
 G\"{o}ran Wendin 
and the Department of
  Microelectronics 
and Nanoscience  at Chalmers University of Technology
 and G\"{o}teborg University for hospitality. The authors also which
 to thank a referee for several constructive remarks.


\end{document}